\documentclass[traditabstract,oldversion]{aa} 

\usepackage{graphicx}
\usepackage{txfonts}

\def\ms{\hbox{\,m\,s$^{-1}$}}         
\def\m2s2{\hbox{\,m$^{2}$\,s$^{-2}$}} 
\def\kms{\hbox{\,km\,s$^{-1}$}}       
\def\vsini{\hbox{$v$\,sin\,$i$}}      

\def\Mjup{\hbox{$\mathrm{M}_{\rm Jup}$}}

\def \1s{$1\,\sigma$}
\def \kid{$\chi^2$}
\def \t0{T$_0$}

\def \cible{HD\,16760}
\def \sophie{{\it SOPHIE}}

\begin{document}
   \title{The $SOPHIE$ northern extrasolar planets\thanks{Based on observations made with 
$SOPHIE$ spectrograph on the 1.93-m telescope at Observatoire de Haute-Provence (CNRS/OAMP), 
France (program 07A.PNP.CONS). }}

   \subtitle{I. A companion close to the planet/brown-dwarf transition around HD\,16760}

\author{
Bouchy, F. \inst{1,2}
\and H\'ebrard, G. \inst{1}
\and Udry, S. \inst{3}
\and Delfosse, X. \inst{4}
\and Boisse, I. \inst{1}
\and Desort, M. \inst{4}
\and Bonfils, X. \inst{4}
\and Eggenberger, A. \inst{4}  
\and Ehrenreich, D. \inst{4}
\and Forveille, T. \inst{4} 
\and Le Coroller, H. \inst{2}
\and Lagrange, A.M., \inst{4}
\and Lovis, C. \inst{3}
\and Moutou, C. \inst{5} 
\and Pepe, F. \inst{3}
\and Perrier, C. \inst{4} 
\and Pont, F. \inst{6}
\and Queloz, D. \inst{3}
\and Santos, N.C. \inst{7}
\and S\'egransan, D. \inst{3}
\and Vidal-Madjar, A. \inst{1}
}

\institute{
Institut d'Astrophysique de Paris, UMR7095 CNRS, Universit\'e Pierre \& Marie Curie, 
98bis Bd Arago, 75014 Paris, France
\and
Observatoire de Haute-Provence, CNRS/OAMP, 04870 St Michel l'Observatoire, France
\and
Observatoire de Gen\`eve, Universit\'e de Gen\`eve, 51 Ch. des Maillettes, 1290 Sauverny, 
Switzerland
\and 
Laboratoire d'Astrophysique, Observatoire de Grenoble, Universit\'e J. Fourier, CNRS, 
BP 53, 38041 Grenoble cedex 9, France
\and 
Laboratoire d'Astrophysique de Marseille, 38 rue Fr\'ed\'eric Joliot-Curie, 
13388 Marseille cedex 13, France
\and
School of Physics, University of Exeter, Exeter, EX4 4QL, UK 
\and
Centro de Astrof\'isica, Universidade do Porto, Rua das Estrelas, 4150-762 Porto, Portugal
}
   \date{Received ; accepted }

 
  \abstract
   {We report on the discovery of a substellar companion or a massive Jupiter orbiting the G5V star HD\,16760 with the 
   spectrograph {\sophie} installed on the OHP 1.93-m telescope. Characteristics and performances 
   of the spectrograph are presented, as well as the {\sophie} exoplanet consortium 
   program. With a minimum mass of 14.3 {\Mjup}, an orbital period of 465 days and an eccentricity of 0.067, 
   HD\,16760b seems to be located just at the end of the mass distribution of giant planets, close to 
   planet/brown-dwarf transition. Its quite circular orbit supports a formation 
   in a gaseous protoplanetary disk.}


\titlerunning{A massive Jupiter around HD\,16760}
\authorrunning{F. Bouchy et al.}

   \maketitle
%

\section{Introduction}

The vast majority of 350 known exoplanets have been found thanks to radial velocity measurements. 
Far to be an old-fashioned technique, Doppler measurements illustrated these last years their capabilities 
to extend the exoplanet search around a large variety of stars. 
The sensitivity of this technique continuously increases, opening the possibility to 
explore the domain of low-mass planets down to few Earth masses, to discover and characterize multiple 
planetary systems, to perform long term surveys to find true Jupiter-like planets, to establish the planetary 
nature and to characterize the transiting candidates of photometric surveys. Doppler surveys for 
exoplanet search require high-precision spectrographs and a significant amount of telescope time 
over a long duration.   

The $SOPHIE$ spectrograph (Bouchy et al. \cite{bouchy06}, Perruchot et al. \cite{perruchot08}) is in operation since October 2006 
at the 1.93-m telescope of Observatoire de Haute-Provence. Benefiting from experience acquired on 
HARPS (Pepe et al. \cite{pepe02}) and taking into account the limitations of the ELODIE spectrograph 
(Baranne et al. \cite{baranne96}), $SOPHIE$ was designed to obtain precise 
radial velocities with much higher throughput than its predecessor and to be operated as a northern 
counterpart of HARPS. This instrument is briefly described in section 2. 
The $SOPHIE$ consortium started in October 2006 a large and comprehensive 
program of search and characterization of exoplanets described in section 3.   
We report in section 4 the detection of a substellar companion or a massive Jupiter 
around HD\,16760 and we discuss in section 5 the properties and nature of this object located 
at the upper limit of the mass distribution of giant planets. 


\section{The {\it SOPHIE} spectrograph}

$SOPHIE$ architecture mainly benefits from ELODIE and HARPS experiences. 
A detailed technical description of this instrument is given by Perruchot 
et al. (\cite{perruchot08}). In this section we briefly describe the main properties 
of the spectrograph and its different observing modes.  
$SOPHIE$ is a cross-dispersed, environmentally stabilized echelle spectrograph 
dedicated to high-precision radial velocity measurements. The detector (EEV-4482) 
is a thinned, back-illuminated, anti-reflection coated 4k x 2k 15-$\mu$m-pixel CCD cooled 
at -100$^o$C, with slow- and fast-readout modes. It records 39 spectral orders covering 
the wavelength domain from 3872 to 6943 {\AA}. The spectrograph is fed through a pair 
of 3''-wide optical fibers for the high-resolution mode (R = 75000, obtained from an extra 
slit), and another pair for the high-efficiency mode (R = 40000, allowing 
one magnitude gain). The high-resolution mode is equipped with a double fiber scrambler 
(Brown \cite{brown90}) to homogenize and stabilize the illumination of the spectrograph entrance. 
For each fiber pair, one aperture is used for starlight whereas the other one, 2' away 
from the first one, can be used either on a Thorium-Argon lamp for tracking spectrograph drift 
({\it thosimult} mode), or on the sky to estimate background pollution, especially in 
case of strong moonlight ({\it objAB} mode). Both aperture can also be simultaneously put 
on Thorium-Argon or tungsten lamps for wavelength or flat-field calibrations, respectively. 
Apart from thermal precautions, the key-point for stability is the encapsulation of the 
dispersive components in a constant pressure tank. This solution stabilizes the air 
refractive index sensitive to atmospheric pressure variations. With such a concept 
typical intrinsic drift of the spectrograph is less than {3 \ms} per hour. The ELODIE 
front-end adaptor (Baranne et al. \cite{baranne96}),  is still used for $SOPHIE$. It holds the calibration 
lamps, the atmospheric dispersion corrector and the guiding system. Compared to ELODIE, 
$SOPHIE$ leads to 1) gain on photon efficiency by a factor of 10 in high-efficiency mode, 
2) increase the spectrograph radial velocity stability by a factor of 3, and 3) increase 
spectral resolution from 42000 to 75000 for the high-resolution mode.  

The spectra are extracted from the detector images and the radial velocities are measured 
online with the $SOPHIE$ pipeline derived and adapted from the HARPS 
one\footnote{http://www.eso.org/sci/facilities/lasilla/instruments/harps/doc/index.html}. 
The spectra extraction includes localization of the 39 
spectral orders on the 2D-images, optimal order extraction, cosmic-ray rejection, wavelength 
calibration and spectral flat-field correction yielding a two dimension spectra (E2DS). 
The orders are then merged and rebinned after correction of the blaze function, yielding a 
one dimension spectra (S1D). The E2DS spectra are cross-correlated with numerical masks 
corresponding to different spectral types (F0, G2, K0, K5, M4); 
the resulting cross-correlation functions (CCFs) are fitted by Gaussians to get the 
radial velocities (Baranne et al. \cite{baranne96}, Pepe et al. \cite{pepe02}). 

Following the approach of Santos et al. (2002), we calibrated the CCF to 
determine the projected rotational velocity {\vsini} and the metallicity index [Fe/H]. 
We also calibrated the CCF to compute the RV photon-noise uncertainty $\sigma_{VR}$. 
Following the approach of Santos et al. (2000), we computed and calibrated 
the chromospheric-activity index $R'_\mathrm{HK}$ based on our {\sophie} spectra.

The $SOPHIE$ radial velocity 
measurements were initially affected by a systematic effect at low signal-to-noise ratio, 
due to CCD charge transfer inefficiency, which increases at low flux level. This effect 
was calibrated and is now corrected by the pipeline, which reallocates the charge lost during 
the readout process on each extracted pixel (Bouchy et al. \cite{bouchy09}). Uncertainties on the radial 
velocity measurements include photon noise, uncertainties in the wavelength calibration, and 
systematic instrumental errors. The photon noise RV uncertainty depends on the signal-to-noise 
of the spectra, as well as on the spectral type and the rotation velocity {\vsini} of the 
observed star. It can be approximated by the semi-empirical estimator 
$ \sigma_\mathrm{RV} = A \times \sqrt{\mathrm{FWHM}} / (\mathrm{S/N} \times C) $, 
where FWHM is the full width at half maximum of the CCF (in same unit as $ \sigma_\mathrm{RV}$), 
$C$ is its contrast (in percent of the continuum),
S/N is the signal-to-noise ratio per pixel at 550~nm,
and the scaling factor $A=1.7$ or 3.4 in high-resolution or high-efficiency mode, respectively. 
For a non-rotating K-dwarf star, a S/N per pixel of 150 provides a photon-noise 
RV uncertainty of 1 \ms. Such a S/N is obtained on 5-mn on a 6.5 magnitude star. The 
uncertainty of the wavelength calibration was estimated to 1 {\ms}. Telescope guiding and 
centering errors in averaged weather conditions are typically of 0.3 - 1 arcsec. In 
high-resolution mode, these errors imply a RV jitter of 3-4 {\ms} due to the insufficient 
scrambling gain of the fiber. This corresponds to the dispersion obtained on the $SOPHIE$ 
measurements around the orbit of HD~189733b, after correction of the stellar jitter 
(Boisse et al. \cite{boisse08}). Uncertainties due to guiding errors are more than twice this level 
in high-efficiency mode due to the absence of scrambler in this instrumental setup.  

The present radial velocity precision obtained on stable stars is about 4-5 {\ms} over 
several semesters. This limitation is mainly due to guiding and centering effects on the 
fiber entrance at the telescope focal plan and the insufficient scrambling provided by 
the fiber and the double scrambler. An upgrade of the Cassegrain fiber adapter is presently 
on-going, including new high-precision guiding camera and new double scrambler, with the goal 
to reach the precision level of 1-2 m/s. 


\section{The {\it SOPHIE} exoplanet program}

The $SOPHIE$ consortium program is devoted exclusively to the study and characterization of 
exoplanets, in continuation of a planet-search program initiated 15 years ago with ELODIE 
spectrograph (Queloz et al. \cite{queloz98}) and in complement to the HARPS program performed in 
the southern hemisphere (Mayor et al. \cite{mayor03}). We started on October 2006 a key program with 
the aims to cover a large part of the exoplanetary science and to bring constraints on the 
formation and evolution processes of planetary systems. Our observing strategies and target 
samples are optimized to achieve a variety of science goals and to solve several important 
issues: 1) mass function of planets below the mass of Saturn, 2) planetary statistical 
properties to constrain the formation and evolution models, 3) relationships between planets 
and the physical and chemical properties of their stars, 4) detection of exoplanets around 
nearby stars, allowing space and ground-based follow-up, 5) deep characterization of known 
transiting exoplanets including long term follow-up and spectroscopic transit analysis. 
All these aspects are treated through 5 complementary sub-programs discussed below and using 
an amount of 60 to 90 nights per semester allocated on $SOPHIE$ at the 1.93-m telescope. \\

{\it\noindent - High precision search for super-Earth}

Only a few percents of the 350 detected planets have masses less than 0.1~M$_{\mathrm{Jup}}$, and due 
to the present precision of radial velocity surveys, the distribution of planetary masses is 
heavily biased against low-mass planets. Recent HARPS discoveries indicate that these low-mass 
exoplanets are not rare and suggest that 30\% of non-active G and K dwarfs solar-type harbor 
Neptune or rocky planets with periods shorter than 50 days (Lovis et al. \cite{lovis08}, Mayor et al. 
\cite{mayor09}). From the ELODIE survey and from our volume-limited sub-program, we pre-selected 
a sample of about 200 non-active bright solar-type stars to explore this domain 
of low-mass planets. \\

{\it\noindent - Giant planets survey on a volume-limited sample}
 
For a large volume-limited sample of 2000 stars, we perform a first screening 
to identify new Hot Jupiters and other Jovian-type planets orbiting 
near and bright stars. Increasing the list of Hot Jupiters offers 
a chance to find a transiting one orbiting a bright star appropriate for additional 
study of planetary atmosphere. This survey will also provide better statistics 
to search for new properties of the distribution of exoplanet parameters. 
We include on this sub-program the long term follow-up and the spin-orbit analysis -- from 
the Rossiter-McLaughlin effect -- of known transiting giant exoplanets to respectively 
detect additional companions and determine the spin-orbit angle of the system.\\ 

{\it\noindent - Search for exoplanets around M-dwarfs}

A systematic search for planets is made for a volume-limited sample of 180 M-dwarfs 
closer than 12 parsecs. Such a survey of  low mass stars will give us a chance 
to derive the frequency of planets as function of the stellar mass. The objectives are 
1) to detect exoplanets of few Earth masses in the habitable zone, 2) to determine the 
statistics of planetary systems orbiting M-dwarfs in combining these 180-M dwarfs sample 
with 100-M dwarfs monitored with HARPS, 3) to identify new potential transiting Hot Neptunes.\\

{\it\noindent - Search for exoplanets around early-type main sequence stars}

A systematic search for planets around a sample of 300 early-type main sequence stars 
(A and F stars) is performed to study the impact of the host star mass on the exoplanet 
formation processes. Such stars were previously not included in the exoplanet surveys due 
to their lack of spectral lines and high rotation broadening. 
A specific pipe-line was developed to compute radial velocity on these specific 
targets (Galland et al. \cite{galland05}) with an accuracy allowing the detection of planets from massive 
hot Jupiters for fast rotating A stars, and down to Neptune-mass planets for the slowest 
F stars. \\

{\it\noindent - Long term follow-up of ELODIE long period candidates}

The ELODIE program for exoplanet search which started on 1994 was performed on a 
sample of 320 G and K stars. About 40 of these stars present evidence of long term trends 
which may be due to giant planets with Jupiter or Saturn like orbit. A long term follow-up 
of these candidates is performed to explore the domain of long period ($\ge$ 10 years) planets.\\

As part of the $SOPHIE$ consortium programs, the detection of four exoplanets have been 
published up to now: HD\,43691b and HD\,132406b (Da Silva et al. \cite{dasilva08}), HD\,45652b (Santos 
et al. \cite{santos08}), and $\theta$~Cygni\,b (Desort et al. \cite{desort09}). These planets have respectively 
minimum masses
of 2.5, 5.6, 0.5 and 2.3 M$_{\rm Jup}$ with a 37, 975, 44 and 154 day periods. 
There were first found from the ELODIE or CORALIE survey then monitored by $SOPHIE$. 
Spectroscopic transits of the massive planets HD 147506b and XO-3b were also observed 
(Loeillet et al. \cite{loeillet08} and H\'ebrard et al. \cite{hebrard08}, respectively), 
allowing a refinement of the parameters 
of the systems, and the detection of a first case of misaligned spin-orbit  for XO-3 (recently 
confirmed by Winn et al. \cite{winn09}). A study of the stellar activity of the transiting planet host star HD\,189733 
is also presented by Boisse et al. (\cite{boisse08}). Recently the transit of the 111-day 
period exoplanet HD\,80606b was established by Moutou et al. (\cite{moutou09}).  

Outside of the consortium programs, $SOPHIE$ plays an efficient role in the 
Doppler follow-up of photometric surveys for planetary transits search. It allowed 
the planetary nature to be established for transiting candidates found by 
SuperWASP (e.g. Collier Cameron et al. \cite{cameron07}, Pollacco et al. \cite{pollacco08}, 
Hebb et al. \cite{hebb08}), by HAT (Bakos et al. \cite{bakos07}) and by CoRoT space mission 
(e.g. Barge et al. \cite{barge08}, Bouchy et al. \cite{bouchy08}, Moutou et al. \cite{moutou08}, 
Deleuil et al. \cite{deleuil08}, Rauer et al. \cite{rauer09}), as well as the parameters of these 
new planets to be characterized, including the measurement of the masses.

In the next section we present the detection of the substellar companion orbiting 
HD\,16760 as part of our sub-program 2 ``{\it Giant planets survey on a volume-limited 
sample}''.

\section{The substellar companion of HD\,16760}

\subsection{Stellar properties of \cible}
\label{section_HD16760_stellar_properties}

\cible\ (HIP\,12638, BD+37\,604) is a G5V star located 50~pc away according 
the Hipparcos parallax. Table~\ref{table_stellar_parameters} summarizes the 
stellar parameters. From spectral analysis of the \sophie\ data using the method 
presented in Santos et al.~(\cite{santos04}),
we derived $T_{\rm eff} = 5608 \pm 20$~K, $\log g = 4.51 \pm 0.10$, 
${\rm [Fe/H]} = -0.02 \pm 0.03$, and $M_* = 0.86 \pm 0.06 \,\rm{M}_{\odot}$, 
which agrees with values from literature. For the temperature and the 
mass, the values we adopt in Table~\ref{table_stellar_parameters} are 
compromise values between ours and those obtained by 
Nordstr\"om et al.~(\cite{nordstrom04}) 
($T_{\rm eff} = 5636$~K and $M_* = 0.91^{+0.08}_{-0.04} \,\rm{M}_{\odot}$). 
We derive  $v\sin i = 2.8 \pm 1.0$~\kms\ from the parameters of the CCF using 
a calibration similar to those presented by Santos et al.~(\cite{santos02}), 
in agreement with the value $v\sin i = 3$~\kms\ from 
Nordstr\"om et al.~(\cite{nordstrom04}). The CCF also allows the value 
${\rm [Fe/H]} = 0.06 \pm 0.05$ to be measured, in agreement but less accurate 
than the metallicity obtained above from spectral analysis.
This target has a quiet chromosphere with no emissions in the \ion{Ca}{ii}
lines ($\log{R'_\mathrm{HK}}=-5.0 \pm 0.1$), making
it a favorable target for planets search from radial velocity measurements.

\begin{table}[h]
\caption{Adopted stellar parameters for \cible.}
  \label{table_stellar_parameters}
\begin{tabular}{lcc}
\hline
\hline
Parameters  & Values & References \\
\hline
$m_v$                		&	$8.744$ 			&	Nordstr\"om et al.~(\cite{nordstrom04}) \\
Spectral~type        		&	G5V				&	Hipparcos catalog \\
$B-V$          			&	$0.71 \pm 0.02$ 	& 	Hipparcos catalog  \\
Distance [pc]     		&	$50 \pm 7$ 		& 	Hipparcos catalog  \\
pmRA [mas/yr]			&	$82.8 \pm 3.1$		& 	Hipparcos catalog  \\
pmDEC [mas/yr]			&	$-110.6 \pm 3.1$	&	Hipparcos catalog  \\
$v\sin i $ [\kms]			&	$2.8 \pm 1.0$		& this work \\
$\log{R'_\mathrm{HK}}$	&	$-5.0 \pm 0.1$		& this work \\
${\rm [Fe/H]}$ 			&	$-0.02 \pm 0.03$	& this work \\
$T_{\rm eff}$ [K]		&	$5620 \pm 30$		& Nordstr\"om et al.~(\cite{nordstrom04}) \& this work \\
$\log g$ [cgi] 			&	$4.51 \pm 0.1$		& this work \\
Mass~$[\rm{M}_{\odot}]$		&	$0.88 \pm 0.08$	& Nordstr\"om et al.~(\cite{nordstrom04}) \& this work \\
\hline
\end{tabular}
\newline
\end{table}

\cible\ has a stellar companion, HIP\,12635 
(Apt~\cite{apt88}, Sinachopoulos~\cite{sinachopoulos07}), 
located $14.562 \pm 0.008$~arcsec in the North and $1.521 \pm 0.002$~mag
fainter. From Hipparcos catalog (Perryman et al., \cite{perryman97}), 
HIP\,12635 has similar distance ($45\pm17$pc) and similar proper motions 
(pmRA=$107\pm17$, pmDEC=$102\pm12$) with \cible, making them 
a likely physical system, with a separation $>700$~AU and an orbital 
period $>10\,000$~years. This would induce tiny radial velocity variations 
on the stars, below $0.2\,$m\,s$^{-1}$yr$^{-1}$.

\subsection{Radial velocity measurements and Keplerian fit}

We acquired with \sophie\ 20 spectra of \cible\ within {\it objAB} mode between December 2006 
and October 2008 under good weather conditions. Two of these 20 spectra were polluted by 
significant Moon contamination. The velocity of the CCF due to the Moon was far enough 
from those of the target to avoid any significant effect on the radial velocity measurement. 
We did not use these two spectra however for the spectral analysis presented in 
\S~\ref{section_HD16760_stellar_properties}.

\begin{figure}[t] 
\begin{center}
\includegraphics[width=8.5cm]{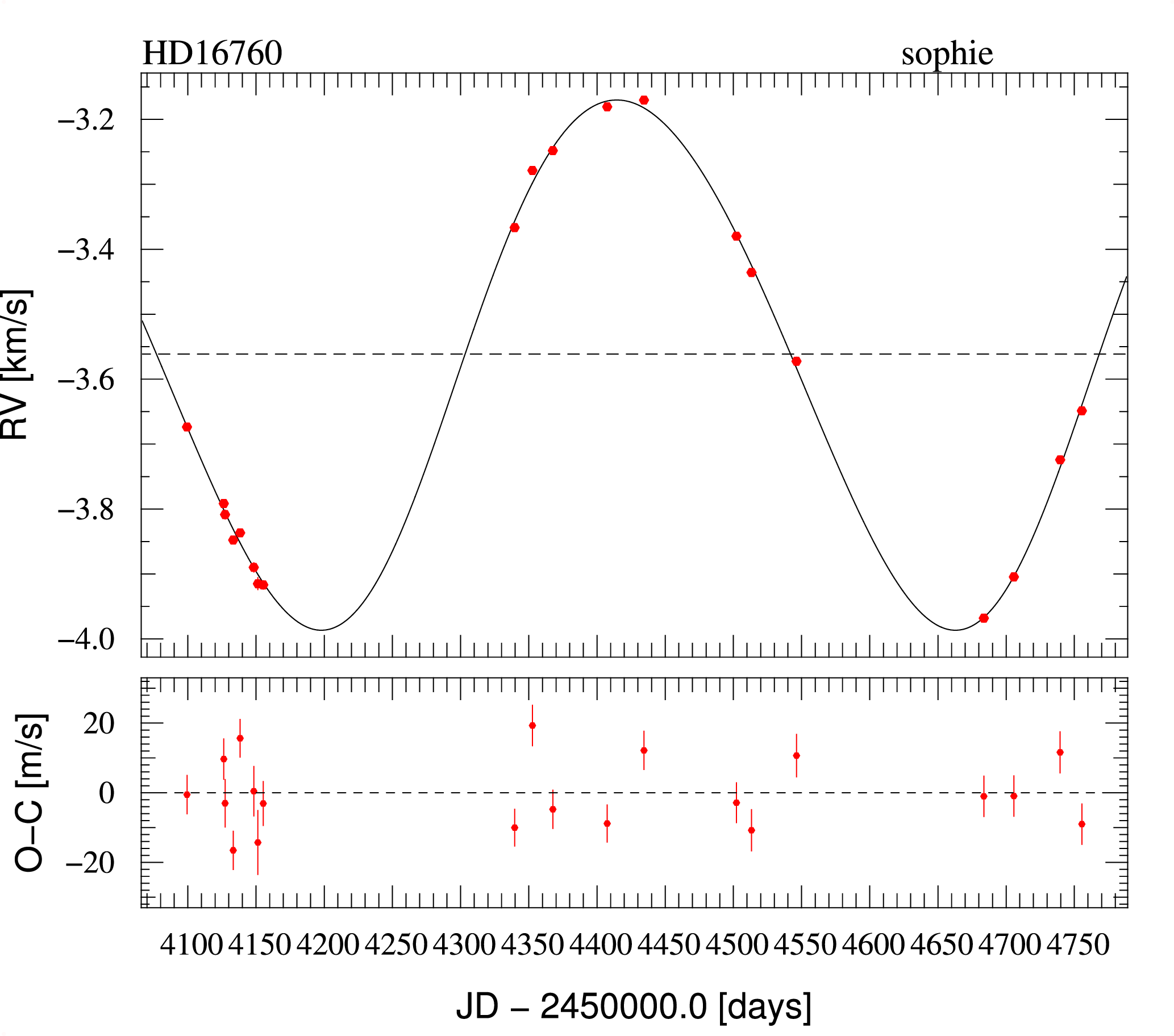}
\caption{\textit{Top:} Radial velocity  {\it SOPHIE} measurements of \cible\ 
as a function of time, and Keplerian fit to the data. 
The orbital parameters corresponding to this 
fit are reported in Table~\ref{table_parameters}. 
\textit{Bottom:} Residuals of the fit with 1-$\sigma$\,Êerror bars.}
\label{fig_omc}
\end{center}
\end{figure}

The derived radial velocities are reported in  Table~\ref{table_rv}, together 
with the journal of the observations. Their typical accuracy is around 6~\ms, 
which is the quadratic sum of three sources of noise: photon noise (3~\ms), 
guiding  (4~\ms), and spectrograph drift (3~\ms). 

\begin{table}[h]
  \centering 
  \caption{Radial velocities of \cible\ measured with {\it SOPHIE}.}
  \label{table_rv}
\begin{tabular}{ccccc}
\hline
\hline
BJD & RV & $\pm$$1\,\sigma$ & exp. time & S/N p. pix. \\
-2\,400\,000 & (km\,s$^{-1}$) & (km\,s$^{-1}$) & (sec) &  (at 550 nm)  \\
\hline
54099.3679  &  -3.6736  &  0.0055  &  300  &  62.5  \\
54126.3439  &  -3.7916  &  0.0058  &  300  &  52.1  \\
54127.2914  &  -3.8085  &  0.0068  &  480  &  36.2  \\
54133.2541  &  -3.8478  &  0.0055  &  224  &  61.4  \\
54138.3414  &  -3.8366  &  0.0054  &  673  &  67.1  \\
54148.3393  &  -3.8898  &  0.0071  &  224  &  35.1  \\
54151.3375  &  -3.9150  &  0.0092  &  225  &  30.0  \\
54155.2737  &  -3.9168  &  0.0063  &  180  &  43.5  \\
54339.6353  &  -3.3666  &  0.0053  &  512  &  78.7  \\
54352.6607  &  -3.2786  &  0.0058  &  380  &  52.0  \\
54367.6375  &  -3.2481  &  0.0055  &  300  &  61.2  \\
54407.4733  &  -3.1806  &  0.0053  &  500  &  79.5  \\
54434.4441  &  -3.1703  &  0.0055  &  620  &  64.2  \\
54502.2489  &  -3.3798  &  0.0057  &  443  &  54.5  \\
54513.2880  &  -3.4358  &  0.0059  &  464  &  53.4  \\
54546.2900  &  -3.5725  &  0.0061  &  1304 & 47.3  \\
54683.6329  &  -3.9680  &  0.0058  &  393  &  51.4  \\
54705.6646  &  -3.9045  &  0.0058  &  379  &  51.2  \\
54739.5899  &  -3.7242  &  0.0059  &  350  &  50.1  \\
54755.5295  &  -3.6488  &  0.0058  &  270  &  51.2  \\
\hline
\end{tabular}
\end{table}

\begin{figure}[h] 
\begin{center}
\includegraphics[width=8.5cm]{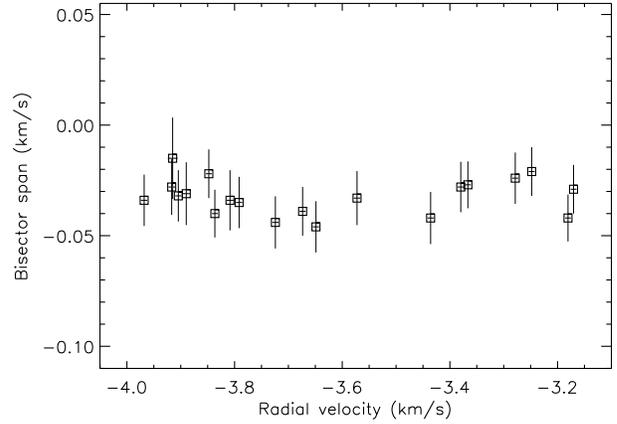}
\caption{Bisector span as a function of the radial velocity.}
\label{fig_bis}
\end{center}
\end{figure}

The radial velocities, shown in Fig.~\ref{fig_omc}, present clear variations of the order 
of hundreds \ms, without significant variations ($\sigma < 10$~\ms) of the CCF bisector 
(Fig.~\ref{fig_bis}), thus in agreement with the reflex motion due to a companion. 
We fitted the data with a Keplerian model.
The solution is a 465-day period oscillation with a semi-amplitude $K=408$~\ms, 
corresponding to a substellar companion, with a minimum mass 
$m_\textrm{p} \sin i   = 14.3$~M$_\mathrm{Jup}$. The derived orbital parameters are 
reported in Table~\ref{table_parameters}, together with error bars, which were 
computed from  $\chi^2$  variations and Monte~Carlo experiments.

The standard deviation of the residuals to the fit is $\sigma(O-C)=10.1$~m\,s$^{-1}$. 
This is higher than the 6-\ms\ estimated uncertainty on the individual measurements. 
Although about 23\% of gaseous giant planets are in a multiple planetary system, 
we do not identify yet an indication for a second body orbiting \cible. 
With a maximum semi-amplitude of 20 \ms, the residuals of the
fit do not exhibit structures, denying a possible inner planet with a
$m_\textrm{p} \sin i \ge$ 0.7 M$_{Jup}$. A longer period planet may induce a drift lower 
than 20~m\,s$^{-1}$yr$^{-1}$ during our observational period.

\begin{table}[h]
  \caption{Fitted orbit and planetary parameters for \cible b.}
  \label{table_parameters}
\begin{tabular}{lcc}
\hline
\hline
Parameters & Values and 1-$\sigma$ error bars & Unit \\
\hline
$V_r$ 				& $-3.561\pm0.004$ 			&   \kms	\\
$P$ 					& $465.1\pm2.3$				&   days 	\\
$e$					& $0.067\pm0.010$				\\
$\omega$ 			& $-128\pm10$					&   $^{\circ}$ \\
$K$					& $408\pm7$					&   \ms	\\
$T_0$ (periastron)		& $2\,454\,723\pm12$			&   BJD 	\\
$\sigma(O-C)$			& 10.1 						&   \ms	\\	
reduced \kid			& 2.0 \\ 
$N_{obs}$					& 20 \\
$m_\textrm{p} \sin i$	 	& $14.3 \pm 0.9$$^\dagger$		&   M$_\mathrm{Jup}$ \\
$a$					& $1.13 \pm 0.03$$^\dagger$		&  AU \\
\hline
$\dagger$: using $M_\star = 0.88\pm0.08$\,M$_\odot$
\end{tabular}
\end{table}



\section{Discussion and Conclusion}

Our RV measurements indicates that a substellar companion with a projected mass 
$m_\textrm{p} \sin i = 14.3$~M$_\mathrm{Jup}$ is orbiting \cible. 
With the degeneracy of inclination angle $i$, it is difficult to conclude about 
the exact nature of this companion. It may correspond to a massive planet, 
formed in a gaseous protoplanetary disk, or a brown dwarf, 
issued from collapse in a giant molecular cloud. 

\begin{figure}[h] 
\begin{center}
\includegraphics[width=8.5cm]{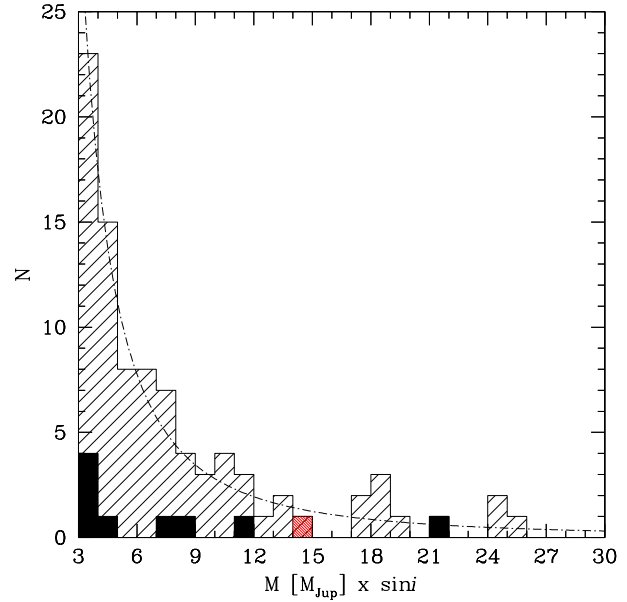}
\caption{Mass distribution of massive planets 
($m_\textrm{p} \sin i \ge$ 3 M$_{\rm Jup}$) and light brown dwarfs ($M_\textrm{c} \sin i 
\le$ 30 M$_{\rm Jup}$) found by radial velocity surveys. The dashed curve corresponds to the 
relation $M^{-2}$ (dN/dM=M$^{-1}$). HD\,16760b, with $m_\textrm{p} \sin i$ = 14.3 \Mjup, is 
identified by the red-filled box. Black-filled symbols correspond to transiting companions.}
\label{histo}
\end{center}
\end{figure}

Figure~\ref{histo} shows the mass distribution of massive planets 
($m_\textrm{p} \sin i \ge$ 3 M$_{\rm Jup}$) and light brown dwarfs ($M_\textrm{c} \sin i 
\le$ 30 M$_{\rm Jup}$) found by radial velocity surveys. From the Extrasolar Planets Encyclopaedia
list\footnote{http://exoplanet.eu}, we completed with HD\,137510b (Endl et al. \cite{endl04}), HD\,180777b 
(Galland et al. \cite{galland06}), and HD\,16760b (this paper), 
totalizing 89 objects including 10 with mass in-between 15 and 30 \Mjup. 
The dashed curve corresponds to the relation $M^{-2}$ (dN/dM=M$^{-1}$). 
The black shaded histogram corresponds to the transiting planets 
with true masses (excluding non-confirming objects SWEEPS-11 and SWEEPS-04). 
It is worthwhile to notice that, although based on 
a small number of object, the mass distribution of transiting planets 
is following the same trend than non-transiting planets. Indeed, the 
ratio of transiting planets over non-transiting planets is about the same 
: 9.2, 9.5 and 10 for the bins 3-6, 6-9 and 9-12 M$_{Jup}$ respectively. 
In this histogram, HD\,16760b seems to be located just at the end of the mass 
distribution of giant planets.
Although based on small numbers, sub-stellar companions 
with minimum mass greater than 17 {\Mjup} do not seem to follow the 
$M^{-2}$ relation.  

\begin{figure}[h] 
\begin{center}
\includegraphics[width=8.5cm]{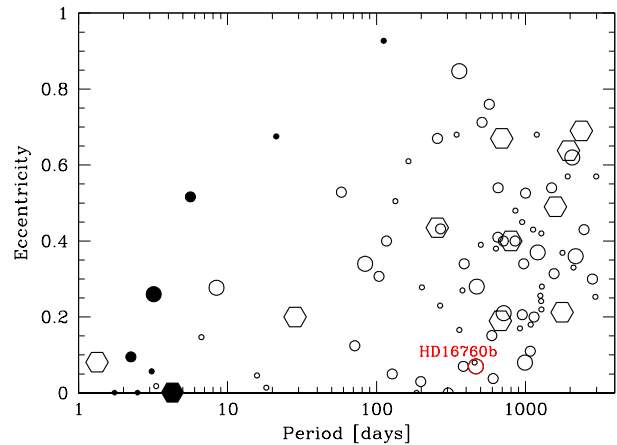}
\caption{Eccentricity - period diagram of massive exoplanets and 
light brown dwarfs. The size of circle is function of the mass (3-5, 5-10, 
10-15 \Mjup). Hexagonal points corresponds to objets with mass greater 
than 15 \Mjup. HD\,16760b, with $P=465$ days and $e=0.067$, is 
identified by the red circle. Black filled symbols correspond to transiting companions.}
\label{ep}
\end{center}
\end{figure}

Figure~\ref{ep} shows the eccentricity - period diagram of massive exoplanets and 
light brown dwarfs. The size of circle is function of the mass (3-5, 5-10, 
10-15 M$_{Jup}$). Hexagonal points corresponds to objets with mass greater 
than 15 M$_{\mathrm{Jup}}$. Black filled symbols correspond to transiting companions. 
{\cible}b confirms the observed trend that more massive
companions are found for longer period planets (Udry \& Santos \cite{udry07}).
We also notice that all companions with mass greater than 15 M$_{\mathrm{Jup}}$ have an 
eccentricity greater than 0.2 except CoRoT-exo-3b (Deleuil et al. \cite{deleuil08}) 
and HD\,41004Bb (Zucker et al. \cite{zucker04}) in very close-in orbit 
(with periods of respectively 4.2 et 1.3 days) tidally circularized. 
The properties of HD\,16760b make it an interresting sub-stellar companion. 
With a mass greater than the Deuterium burning limit (13 \Mjup), it may be 
defined as a brow-dwarf. However its quite circular orbit supports a formation 
in a gaseous protoplanetary disk. 
On another way, Halbwachs et al. (\cite{halbwachs05}) studied the eccentricity
distribution for exoplanets and binary stars with a mass ratio smaller than
0.8 (non twin binaries). They found that exoplanets have orbits with 
eccentricities significantly smaller than those of the non-twin binaries, 
reinforcing the hypothesis that planetary systems and stellar binaries are not the
products of the same physical process. 

{\cible}b is in a visual and physical binary system. However, in the mass-period and 
eccentricity-period diagrams, HD\,16760\,b is located in a region not much 
populated by planets in binary systems. The discovery of
this long-period low-eccentricity planet thus adds to the growing evidence
that contrary to short-period planets, long-period ($\gtrsim$100 days) planets
residing in binaries possess the same statistical properties as their
counterparts orbiting single stars (Eggenberger et al. \cite{Eggenberger04}, 
Mugrauer et al. \cite{Mugrauer05}, Desidera et al. \cite{Desidera07}).

{\cible}b would induce a motion of its host star of at least 
$\pm 0.35$ milli-arcsec. The future Gaia ESA space mission 
scheduled for launch in late-2011, should be able to detect this system 
from astrometry, and thus would allow the inclination of the system to be measured 
and the true mass to be determined. Detailed caracterization of this sub-stellar companion 
close to planet/brow-dwarf transition will help to distinguish the differences of 
formation processes between these two populations.

\begin{acknowledgements}
The authors thanks all the staff of Haute-Provence Observatory for their 
contribution to the success of the {\sophie} project and their support 
at the 1.93-m telescope. 
We wish to thank the ``Programme National de Plan\'etologie'' (PNP) of CNRS/INSU, the 
Swiss National Science Foundation, and the French National Research Agency (ANR-08-JCJC-0102-01 and 
ANR-NT05-4-44463) for their continuous support to our planet-search 
programs. F.B. acknowledges S.F.Y.B.L.S for continuous support and advices.

\end{acknowledgements}

%
%
%
%
%


\begin{thebibliography}{}

\bibitem[1988]{apt88} 
Abt, H. A. 1988, \apj, 331, 922

\bibitem[2007]{bakos07} 
Bakos, G. A., Shporer, A., Pal, A., et al. 2007, \apj, 671, L173

\bibitem[1996]{baranne96} 
Baranne, A., Queloz, D., Mayor, M., et al. 1994, \aaps, 119, 373

\bibitem[2008]{barge08} 
Barge, P., Baglin, A., Auvergne, M., et al. 2008, \aap, 482, L17

\bibitem[2009]{boisse08} 
Boisse, I., Moutou, C., Vidal-Madjar, A., et al. 2009, \aap, 495, 959

\bibitem[2006]{bouchy06} 
Bouchy, F., and the Sophie team, 2006, in \textit{Tenth Anniversary of 51~Peg-b}: 
Status of and prospects for hot Jupiter studies, 
eds. L.~Arnold, F.~Bouchy \& C.~Moutou, 319

\bibitem[2008]{bouchy08} 
Bouchy, F., Queloz, D., Deleuil, M., et al. 2008, \aap, 482, L25

\bibitem[2009]{bouchy09} 
Bouchy, F., Isambert, J., Lovis, C., et al. 2009, 
in \textit{Astrophysics Detector Workshop}, EAS Publications Series, 37, 247

\bibitem[1990]{brown90}
Brown, T.M., 1990, in: CCDs in Astronomy, Jacoby G. (ed.), San Francisco, ASP, ASP Conf. Ser. 8, 335

\bibitem[2007]{cameron07} 
Collier Cameron, A., Bouchy, F., H\'ebrard, G., et al. 2007, \mnras, 375, 951

\bibitem[2006]{dasilva06} 
Da Silva, R., Udry, S., Bouchy, F., et al. 2006, \aap, 446, 717

\bibitem[2008]{dasilva08} 
Da Silva, R., Udry, S., Bouchy, F., et al. 2008, \aap, 473, 323

\bibitem[2008]{deleuil08}
Deleuil, M., Deeg, H., Alonso, R., et al., 2008, \aap, 491, 889 

\bibitem[2007]{Desidera07}
Desidera, A \& Barbieri, M., 2007, \aap, 462, 345

\bibitem[2009]{desort09}
Desort, M., Lagrange, A.-M., Galland, F., et al., 2009, \aap, in press

\bibitem[2004]{endl04}
Endl, M., Hatzes, A., Cochran, W.D., et al., 2004, \apj, 611, 1121

\bibitem[2004]{Eggenberger04}
Eggenberger, A., Udry, S., \& Mayor, 2004, \aap, 417, 353

\bibitem[2005a]{galland05}
Galland, F., Lagrange, A.-M., Udry, S., et al., 2005a, \aap, 443, 337

\bibitem[2005b]{galland06}
Galland, F., Lagrange, A.-M., Udry, S., et al., 2005b, \aap, 444, L21

\bibitem[2005]{halbwachs05}
Halbwachs, J.L., Mayor, M., \& Udry, S., 2005, \aap, 431, 1129

\bibitem[2008]{hebb08} 
Hebb, L., Collier-Cameron, A., Loeillet, B., et al. 2008, \apj, 693, 1920

\bibitem[2008]{hebrard08} 
H\'ebrard, G., Bouchy, F., Pont, F., et al. 2008, \aap, 481, 52

\bibitem[2008]{loeillet08} 
Loeillet, B., Shporer, A., Bouchy, F., et al. 2008, \aap, 481, 529

\bibitem[2009]{lovis08}
Lovis, C., Mayor, M., Bouchy, F., et al., 2009, in: \textit{Transiting PLanets}, 
Proceedings of the IAU Symposium, vol. 253, 502

\bibitem[2003]{mayor03}
Mayor, M., Pepe, F., Queloz, D., et al., 2003, The Messenger, 114, 20

\bibitem[2009]{mayor09}
Mayor, M., Udry, S., Lovis, C., et al., 2009, \aap, 493, 639 

\bibitem[2008]{moutou08}
Moutou, C., Bruntt, H., Guillot, T., et al., 2008, \aap, 488, 47

\bibitem[2009]{moutou09}
Moutou, C., H\'ebrard, G., Bouchy, F., et al., 2009, \aap, 498, L5

\bibitem[2005]{Mugrauer05}
Mugauer, M., Neuh\"auser, R, Seifahrt, A., et al., 2005, \aap, 440, 1051

\bibitem[2004]{nordstrom04} 
Nordstr\"om, B., Mayor, M., Andersen, J., Holmberg, J., Pont, F., J\o rgensen, B.~R., Olsen, E.~H., Udry, S., Mowlavi, N.	
2004, \aap, 418, 989

\bibitem[2002]{pepe02} 
Pepe, F., Mayor, M., Galland, F., et al. 2002, \aap, 388, 632

\bibitem[2008]{perruchot08}
Perruchot, S., Kohler, D., Bouchy, F., et al., 2008, in \textit{Ground-based and Airborn
Instrumentation for Astronomy II}, Edited by McLean, I.S., Casali, M.M., Proceedings of the 
SPIE, vol. 7014, 70140J

\bibitem[1997]{perryman97}
Perryman, M.A.C., Lindegren, L., Kovalevsky, J., et al., 1997, \aap, 323, L49

\bibitem[2008]{pollacco08} 
Pollacco, D., Skillen, I., Collier Cameron, A., et al. 2007, \mnras, 385, 1576

\bibitem[1998]{queloz98}
Queloz, D., Mayor, M., Sivan, J.P., et al., 1998, in \textit{Brown dwarfs and extrasolar 
planets}, Edited by Rebolo, R., Martin, E.L., Osorio, M.R.Z., ASP Conf. Ser. 134, 324

\bibitem[2009]{rauer09}
Rauer, H., Queloz, D., Csizmadia, Sz., et al., 2009, \aap, in press 

\bibitem[2000]{santos00} 
Santos, N. C., Mayor, M., Naef, D., et al. 2000,  \aap, 361, 265

\bibitem[2002]{santos02} 
Santos, N. C., Mayor, M., Naef, D., et al. 2002, \aap, 392, 215

\bibitem[2004]{santos04} 
Santos, N. C., Israelia, G., Mayor, M. 2004, \aap, 415, 1153

\bibitem[2008]{santos08} 
Santos, N. C., Udry,  S., Bouchy, F., et al. 2008, \aap, 487, 369

\bibitem[2007]{sinachopoulos07} 
Sinachopoulos D., Gavras P., Dionatos O., Ducourant C., Medupe T. 2007, \aap, 472, 1055

\bibitem[2007]{udry07}
Udry, S., \& Santos, N.C., 2007, \araa, 45, 397

\bibitem[2009]{winn09}
Winn, J. N., Johnson, J.A, Fabrycky, D., et al., 2009, \apj, 700, 302 

\bibitem[2004]{zucker04}
Zucker, S., Mazeh, T., Santos, N.C., et al., 2004, \aap, 426, 695

\end{thebibliography}
\end{document}